# Deep Convolutional Generative Modeling for Artificial Microstructure Development of Aluminum-Silicon Alloy

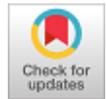

Akshansh Mishra, Tarushi Pathak

*Abstract: Machine learning which is a sub-domain of an Artificial Intelligence which is finding various applications in manufacturing and material science sectors. In the present study, Deep Generative Modeling which a type of unsupervised machine learning technique has been adapted for the constructing the artificial microstructure of Aluminium-Silicon alloy. Deep Generative Adversarial Networks has been used for developing the artificial microstructure of the given microstructure image dataset. The results obtained showed that the developed models had learnt to replicate the lining near the certain images of the microstructures.*

*Keywords: Machine Learning; Deep Generative Modeling; Artificial Microstructure; Generative Adversarial Network*

## I. INTRODUCTION

Machine Learning (ML) is a subset of artificial intelligence (AI). It is a computational science technique which focuses on interpreting and analysing patterns which gives the ability of decision making beyond the human interaction. Machine learning can be mainly classified into two types i.e. supervised learning and unsupervised learning [1-4]. In supervised machine learning setup we have data (x, y) where x is data and y is label. The main objective is to learn and construct a function that is mapping from our data x to label y i.e. x → y and this label can take other various forms. For example, classification, regression, object detection, semantic segmentation etc. In unsupervised machine learning we have unlabelled data and the main objective is to learn some underlying hidden structure of the data and it is relatively unexplored research area. The examples of unsupervised learning are feature learning, clustering, density estimation, dimensionality reduction etc.

The present work utilizes generative modelling which are class of models for unsupervised machine learning technique in which for a given training data our objective is to generate new samples from some distribution and $P_{model}(x)$ is learned to generate images from the same distribution i.e. $P_{data}(x)$ which is shown in Fig.1.

Generative model addresses density estimation which is a core problem in unsupervised learning. Generative models is used for generating samples for artwork, super-resolution, colorization etc. and also generative models of time series data can be used for simulation and planning (reinforcement learning applications) [5-8]. Generative models are classified into two types i.e. explicit density estimation where we are going to define and solve for $P_{model}(x)$ and implicit density estimation where a learned model can sample from $P_{model}(x)$ without explicitly defining it.

Generative modeling is completely different from discriminative modeling. Generative models try to learn to make a realistic representation of some classes. GAN learns by making a discriminator and a generator compete against each other. GAN consists of two components i.e. Generator and discriminator which are typically two different neural networks. The job of a generator is to learn and further generate fakes that look real in order to fool the discriminator while on the other hand, the duty of the discriminator is to learn and further distinguish between the real and fakes data. The generator forges fake images to try to look as realistic as possible and it does this in the hopes of fooling the discriminator. So basically in order to start these whole processes of fooling and identifying, we need a collection of real images. At the beginning of the process, the generator is actually not very sophisticated i.e. it doesn't have the knowledge of creating a real set of images. The generator is not allowed to look into real images while the elementary discriminator is allowed to look into images but the irony is that it has no idea that which data is real and which data is fake. So both models learn from their competition with each other until the images produced by the generator are good enough to fool the discriminator.

Lee et al. [9] carried out a case study in fault diagnosis and detection by using generative adversarial neural networks. The results indicated that under the given condition the oversampling by generative adversarial networks performed very well. Luo et al. [10] carried out a case study for machine fault diagnosis by using conditional deep convolutional generative adversarial networks. It was concluded that the proposed method can deal with the fault diagnosis objective more effectively. Mao et al. [11] used a generative adversarial network for designing complex architecture materials as shown in Fig.2. Singh et al. [12] proposed generative adversarial networks model for synthetic defect generation in assembly and test manufacturing.

**Deep Convolutional Generative Modeling for Artificial Microstructure Development of Aluminum-Silicon Alloy**

The results generated realistic-looking semiconductor defects images of various types i.e. scratches, misplaced epoxy, foreign materials, and die-chipping defects.

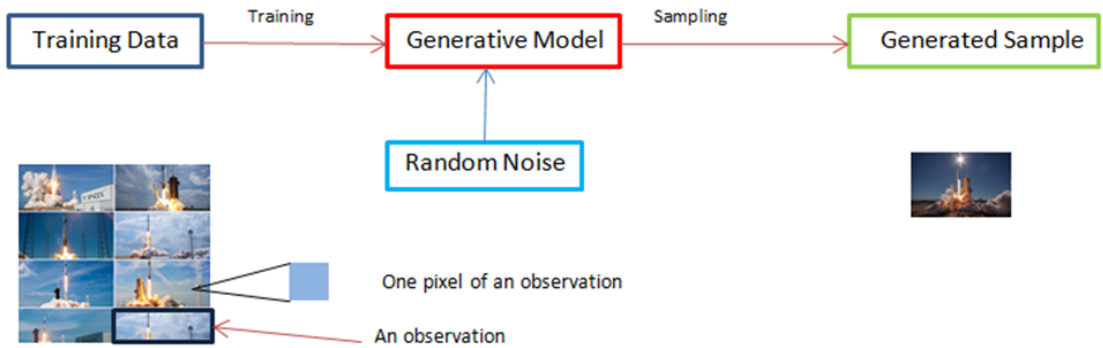

Fig 1. Working architecture of GAN

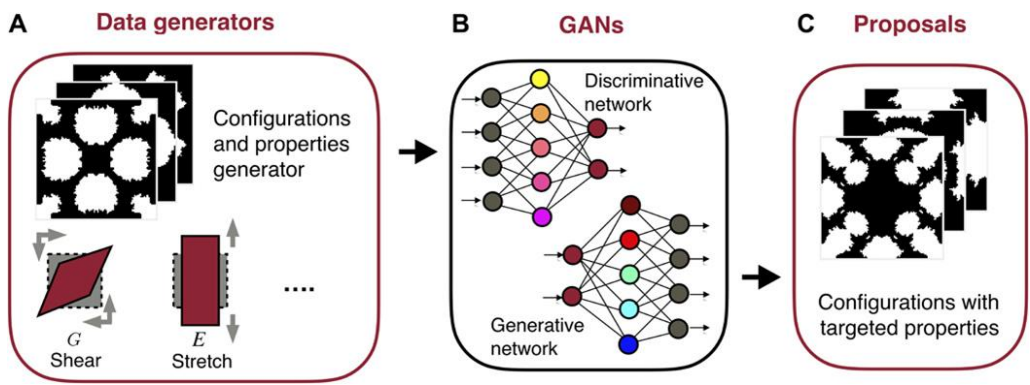

Fig 2. Designing of complex architecture by using generative adversarial networks [11]

## II. MATERIAL AND METHODS

**A.** Initially, the dataset consisted of 3 images, which is far too less for any deep learning task. So, more images were taken off the internet. This increased the size of the dataset to 67 images, which was still too less. Using data augmentation, the dataset was increased from 67 images to 13000 images. This was done by iterating each image from the dataset through the Keras' ImageDataGenerator, where they underwent width shift, height shift, shear range, zoom range, horizontal flip, and vertical flip, followed by saving them. Hence, the images used in this experiment are augmented, because of which they might have lines around them or white spaces. The inspiration of GANs is having two deep learning models pitted against each other. These two models are called Generator and Discriminator. The Generator is first fed a noise input through which it tries to replicate the original data distribution such that the Discriminator, which is in charge of classifying whether the image formed is from the training set or not, classifies it as a real image. Mathematically speaking, the Discriminator model is trained to maximize the probability of correctly classifying both training examples and samples from the Generator, whereas the Generator is trained to minimize $log(1-D(G(z)))$, where z is the noise input, $G(z)$ is the image generated and $D(G(z))$ is the probability of the image being a real image. Discriminator and Generator play the following two-player minimax game with value function V(G,D):

$$V(D,G)=E_{x\sim pdata(x)}[logD(x)]+E_{z\sim pz(z)}[log(1-D(G(z)))] \quad (1)$$

Equation (1) is the loss function for our GAN. The class of GANs used for this experiment is the Deep Convolutional General Adversarial Net or the DCGAN.

## III. RESULTS AND DISCUSSION

Figure 3 is a model summary of the Generator used. If we were to visualize it would look somewhat like the second image i.e. Figure 4. If the above result is compared with typical convolutional models we will notice that it is missing the fully connected layers and the pooling layers an ideal convolutional network is supposed to have. The fully connected layers were nullified in order to allow the Discriminator and the Generator to be directly connected. If we look at the model summary, we can see that the convolutions being used are stridden and are 2D Transpose Convolutions. This is done to let the convolutions learn how to upscale themselves.

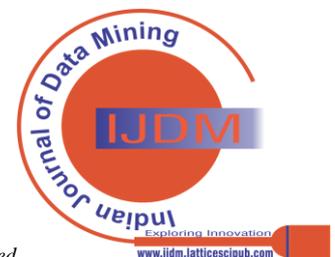

The images fed into this convolutional network instead of being scaled from 0 to 1, are scaled from -1 to 1 as the output function is tanh. The weights are initialized to a normal distribution. Batch Normalization layers were used with mean =0 and standard deviation set to 0.2. The Batch Normalization layers were not added to the output layer in order to maintain the stability. The activation function used is the RELU and the final layer is tanh. The optimizer used is Adam optimizer with the beta as 0.5.

Fig 5 is a model summary of the discriminator model. If we visualize it would look somewhat like Fig 6. A discriminator's job is to classify whether the image is from a training dataset or a generated image. In simpler words, it is a classification model. Much like the Generator, the fully connected layers and the pooling layers are absent. The convolutions are stridden in order to enable the discriminator to learn how to downscale itself. A classification convolutional model is generally accompanied with fully connected layers at the end but in this case, you will notice just a flat convolution towards the end. The sigmoid activation function is simply applied to this flat layer to obtain the probability. These layers also have Batch Normalization with mean=0 and standard deviation=0.2. They are also initialized with weights of normal distribution. The activation functions used are LeakyRelu and sigmoid for the last layer. The GAN training happens in alternate periods, that is, first the discriminator trains for one or two epochs, then the generator trains for one or two epochs. The GAN converges when it reduces the Discriminator's accuracy to 50%, which implies that the Discriminator is flipping a coin to make a decision. The learning rate experimented between 0.0002 and 0.0005. The model seemed to learn faster at the learning rate of 0.0005, so it was decided as the final learning rate. The image indicated by Figure 7 is the generated images and the ones in Figure 8 are training images. As can be observed, the model seems to have learned to replicate the lining near-certain images.

Figure 9 shows the performance of the Generator and Discriminator during the training. From the images and the graph, it is easy to interpret that the model can perform much better with more training and more data.

```
Generator(
  (main): Sequential(
    (0): ConvTranspose2d(1000, 512, kernel_size=(4, 4), stride=(1, 1), bias=False)
    (1): BatchNorm2d(512, eps=1e-05, momentum=0.1, affine=True, track_running_stats=True)
    (2): ReLU(inplace=True)
    (3): ConvTranspose2d(512, 256, kernel_size=(4, 4), stride=(2, 2), padding=(1, 1), bias=False)
    (4): BatchNorm2d(256, eps=1e-05, momentum=0.1, affine=True, track_running_stats=True)
    (5): ReLU(inplace=True)
    (6): ConvTranspose2d(256, 128, kernel_size=(4, 4), stride=(2, 2), padding=(1, 1), bias=False)
    (7): BatchNorm2d(128, eps=1e-05, momentum=0.1, affine=True, track_running_stats=True)
    (8): ReLU(inplace=True)
    (9): ConvTranspose2d(128, 64, kernel_size=(4, 4), stride=(2, 2), padding=(1, 1), bias=False)
    (10): BatchNorm2d(64, eps=1e-05, momentum=0.1, affine=True, track_running_stats=True)
    (11): ReLU(inplace=True)
    (12): ConvTranspose2d(64, 3, kernel_size=(4, 4), stride=(2, 2), padding=(1, 1), bias=False)
    (13): Tanh()
  )
)
```

**Fig 3. Model Summary of Generator**

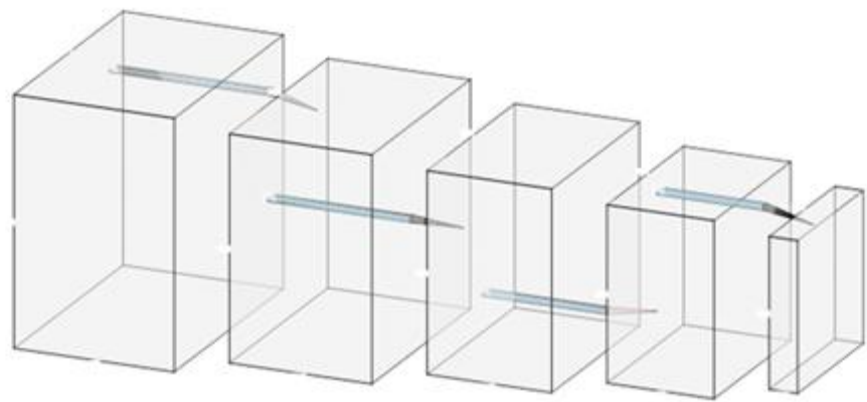

**Fig 4. Visualization of Generator**

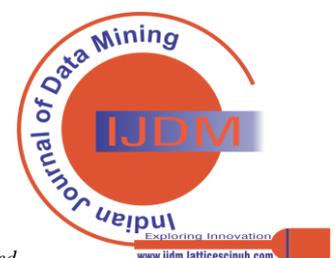

**Deep Convolutional Generative Modeling for Artificial Microstructure Development of Aluminum-Silicon Alloy**

```
Discriminator(
  (main): Sequential(
    (0): Conv2d(3, 64, kernel_size=(4, 4), stride=(2, 2), padding=(1, 1), bias=False)
    (1): LeakyReLU(negative_slope=0.2, inplace=True)
    (2): Conv2d(64, 128, kernel_size=(4, 4), stride=(2, 2), padding=(1, 1), bias=False)
    (3): BatchNorm2d(128, eps=1e-05, momentum=0.1, affine=True, track_running_stats=True)
    (4): LeakyReLU(negative_slope=0.2, inplace=True)
    (5): Conv2d(128, 256, kernel_size=(4, 4), stride=(2, 2), padding=(1, 1), bias=False)
    (6): BatchNorm2d(256, eps=1e-05, momentum=0.1, affine=True, track_running_stats=True)
    (7): LeakyReLU(negative_slope=0.2, inplace=True)
    (8): Conv2d(256, 512, kernel_size=(4, 4), stride=(2, 2), padding=(1, 1), bias=False)
    (9): BatchNorm2d(512, eps=1e-05, momentum=0.1, affine=True, track_running_stats=True)
    (10): LeakyReLU(negative_slope=0.2, inplace=True)
    (11): Conv2d(512, 1, kernel_size=(4, 4), stride=(1, 1), bias=False)
    (12): Sigmoid()
  )
)
```

**Fig 5. Model Summary of Discriminator**

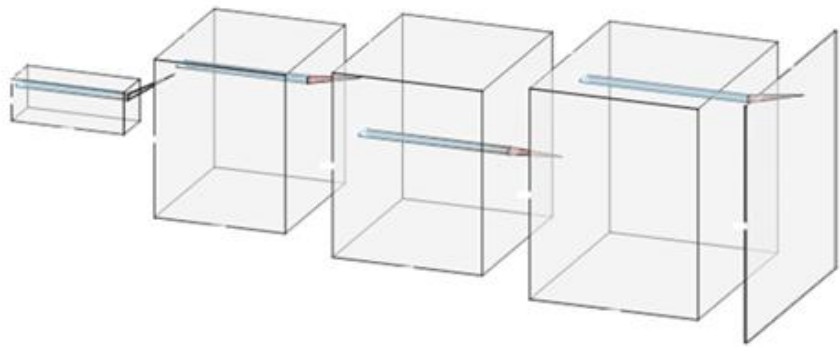

**Fig 6. Visualization of Discriminator Model**

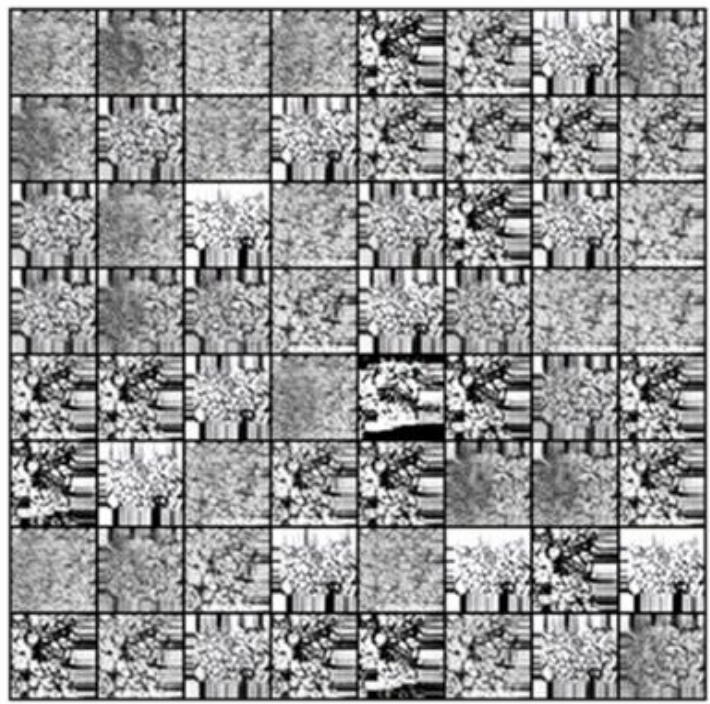

**Fig 7. Artificial generated microstructure of Aluminum-Silicon Alloy**

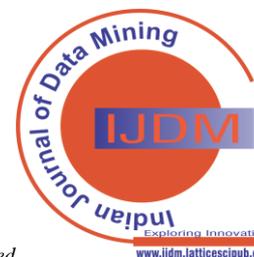

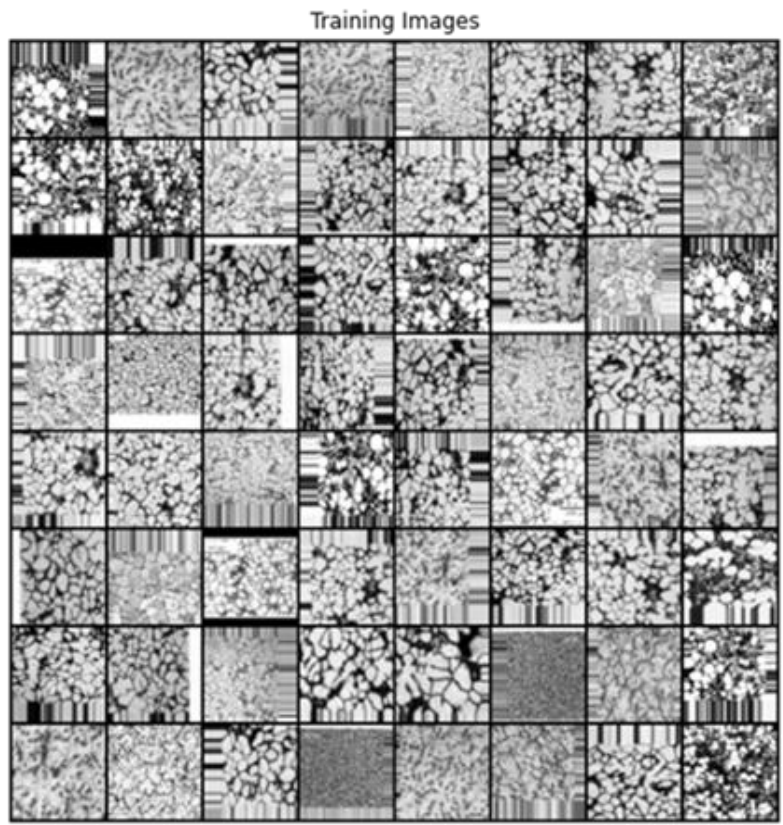

**Fig 8. Training images used to generate artificial images**

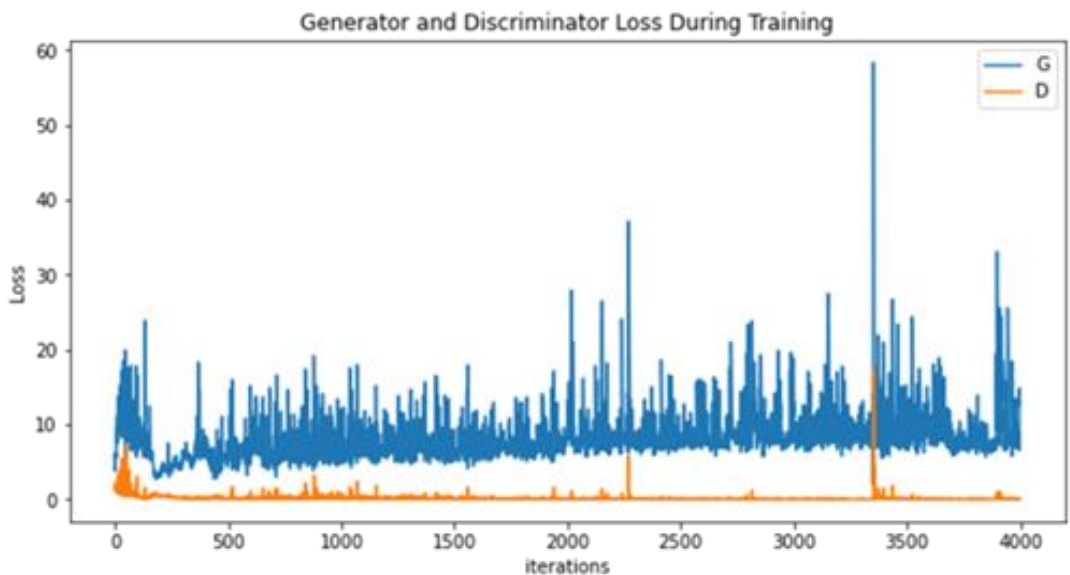

**Fig 9. Plot of loss with respect to number of iterations**

## IV. CONCLUSION

It is observed that the Machine Learning algorithm i.e. Deep Convolutional Generative Adversarial Network is successfully implemented on the available microstructure dataset of Aluminum-Silicon alloy. It is also concluded that the developed model is able to learn from the set of training data and it further had developed the ability to generate a new data with same characteristics as the training data.

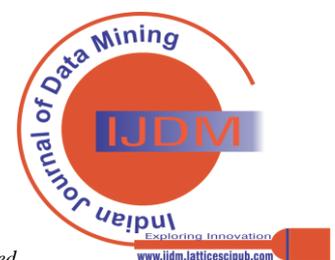

## AUTHORS PROFILE

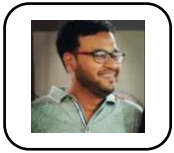

**Akshansh Mishra,** works on the application of Artificial Intelligence in manufacturing process. He is currently working as a Research and Development Scientist at Neural Net and Stir Research Technologies.

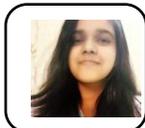

**Tarushi Pathak,** is a computer science undergraduate at SRM Institute of Science and Technology, Kattankulathur, Chennai. She has registered herself as one of the top 25 contributors at Student Code-In, Open Source Event. Her primary field of interest is Deep Learning and Data Science.